# Helping HPC Users Specify Job Memory Requirements via Machine Learning

Eduardo R. Rodrigues[1], Renato L. F. Cunha[1], Marco A. S. Netto[1], Michael Spriggs[2]

[1]IBM Research
[2]IBM Systems

*Abstract*—Resource allocation in High Performance Computing (HPC) settings is still not easy for end-users due to the wide variety of application and environment configuration options. Users have difficulties to estimate the number of processors and amount of memory required by their jobs, select the queue and partition, and estimate when job output will be available to plan for next experiments. Apart from wasting infrastructure resources by making wrong allocation decisions, overall user response time can also be negatively impacted. Techniques that exploit batch scheduler systems to predict waiting time and runtime of user jobs have already been proposed. However, we observed that such techniques are not suitable for predicting job memory usage. In this paper we introduce a tool to help users predict their memory requirements using machine learning. We describe the integration of the tool with a batch scheduler system, discuss how batch scheduler log data can be exploited to generate memory usage predictions through machine learning, and present results of two production systems containing thousands of jobs.

## I. INTRODUCTION

Batch schedulers such as IBM Spectrum LSF, TORQUE, PBS, among others rely on users specifying job requirements. Examples of user input are amount of memory, number of processors, and requested time. These values are hard for the user to specify due to various options of job and environment configurations and intricate effects of these values on overall job response time [1, 2]. Job specifications and their impact on scheduling decisions have been vastly investigated by several research groups [3, 4, 5, 6].

Parameters such as amount of memory and number of processors, for instance, impact the performance of the user application but also scheduler dynamics and overall cluster utilization. If users ask for a certain amount of memory but use much less, resources are wasted and other users cannot benefit of early cluster access. Estimations on how long jobs will wait in a queue and their execution time have a direct impact on users' planning [7].

Systems such as XSEDE[1] utilize prediction techniques for queue waiting time. From our experiments, we observed that applying those techniques to predict job memory requirement does not produce the best predictions. In this paper we introduce a tool to assist users predicting memory usage using batch scheduler logs—the tool could also be used to predict other features related to resource allocation in HPC environments. The tool is based on machine learning techniques

Author preprint. To appear in the Proceedings of the Third Annual Workshop on HPC User Support Tools (IEEE).

[1]XSEDE - https://www.xsede.org/

and is invoked through command line interface, and can be extended to support the existing LSF GUI (IBM Spectrum LSF Application Center). The contributions of the paper are therefore:

- Machine learning based tool to predict resource allocation features for HPC environments, such as job memory usage (§ II);
- Evaluation of the tool to understand how different machine learning techniques perform in the task of predicting memory usage (§ III).

We describe in details the design and implementation of the tool and evaluated it in a variety of scenarios with workloads from real HPC settings.

## II. PREDICTION TOOL

In this section, we describe the architecture of the tool and the machine learning techniques used. We particularly describe how the data is obtained and what preprocessing is required. We also present the training and validation scheme we created and the voting system we devised to combine the predictions of top techniques in a single predictor.

### A. System Description

Figure 1 illustrates the data flow in our tool and the three main components: data collection and conversion, model construction, and predictions. The first step in order to make predictions is to collect historical data. Our tool interfaces directly with the LSF batch scheduler to collect relevant data to predict run time, wait time and memory usage. In this paper, we only discuss memory usage predictions though.

There are two modes of collection: on-line and off-line. In the on-line mode, the tool collects historical data while the jobs are submitted to the cluster. It pulls periodically the relevant data from LSF's memory using the public LSF API. In the off-line mode, the tool collects data from the databases that LSF stores for its own statistics. This last mode is suitable for exploiting the data the users have already accumulated throughout the history of the system. This mode is also important while testing the tool: it allows a fast way to generate data to be used by the predictors. However, there are some statistics that LSF does not persist and can only be collected in the on-line mode.

Two types of data are collected: features and labels. Features describe the conditions under which an event happened, and labels describe what happened, in a dimension of interest.

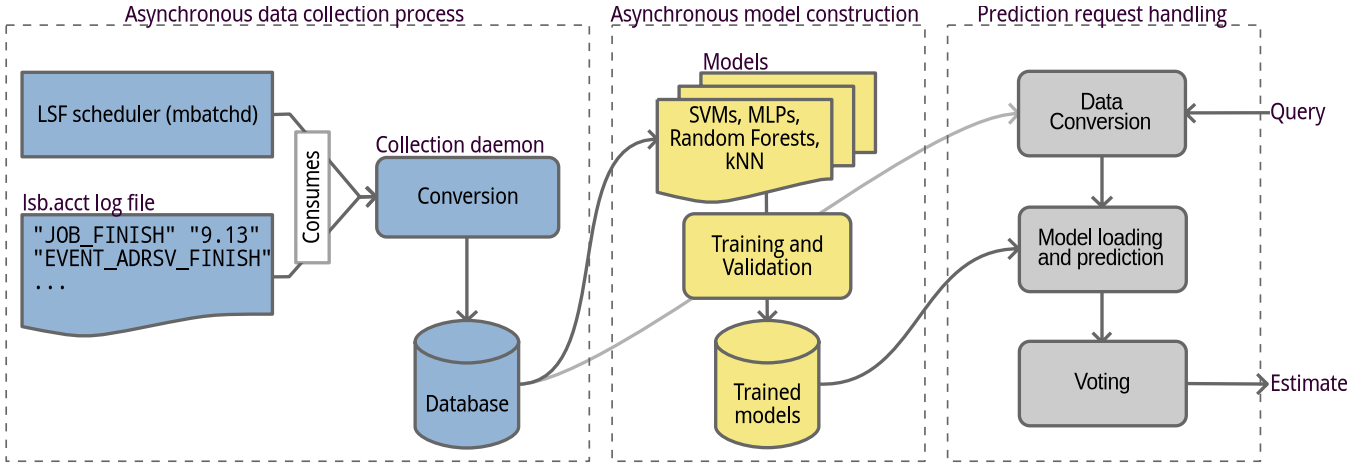

Fig. 1. High-level view of the tool's operation with three main components. The component in the left handles data capture and persistence and runs as a daemon. The middle component loads the persisted data and trains models asynchronously. The component in the right is the interface with the world and handles user requests.

Specifically, our features correspond to characteristics of the jobs, such as user id, number of requested processors, working directory, and so on. All of which are known data at job submission time. The label is the amount of memory effectively used and is only available after the job finishes.

In our tool we collect a wide range of features. We pre-process these features and store them in a separate embedded database. This pre-processing stage transforms the raw features into a format that is more suitable for the machine learning techniques we use. For example, we transform text features, such as working directory path, into a numeric feature (a foreign key to a features table). In addition to that, we augment the data by including some derived features, such as submission weekday.

Even though many features are collected, not all of them need to be always used. We implemented a configurable selector that allows the user to experiment with different sets of features. In Figure 1 this is represented by the arrow connecting the SQLite database in the data collection process to the data conversion component in the prediction request handler. Being written in this way, in case new features should be used to predict a given label, the user only needs to update the row related to that label in the database. One could perform feature selection [8] for determining the best subset of features to be used for each label. The features we used in our experiments are shown in Table I.

The label is the maximum memory used. However, instead of treating the prediction as a regression, we placed the label into bins and transformed the prediction into a classification problem. In addition to simplifying the evaluation, one reason for discretizing memory usage is that it helps scheduling performance. If we use the discretized memory usage predictions directly from the tool, then the scheduler has fewer unique resource requirements to deal with. In our experiments we used bins of size 512MiB.

During prediction, the tool obtains features and labels of a set of jobs from the database. This set is a sliding window that contains the most recent jobs in the database. The size of the window is adjustable, and, in principle, could be optimized, i.e. the size of the window could be adjusted to improve prediction. In our experiments, however, we used a fixed window of 10,000 entries. In this step, the tool also creates dummy variables for the categorical variables by using one-hot encoding [9]. Moreover, for some machine learning techniques, this step also normalizes the numerical features, so that features with high variability do not have more influence in the prediction.

The features and labels obtained in the previous step are divided in two subsets. One for training and the other for validation. The training set is composed of 9,000 entries and the remaining 1,000 entries are used to validate the trained models. Each one of the methods we describe in the next section is trained with the training set and evaluated with the validation set. The metric used to evaluate is accuracy, i.e. the percentage of correctly classified samples. Each trained model is then persisted to disk so that the prediction step can use them asynchronously. The computed accuracy in the validation set is used to sort the models from best performing to the worst performing and is also used as a weight in the method described below.

In order to issue the final prediction, the tool performs a poll of the top predicting methods, ranked by the validation accuracy. In our experiments, we use the top four, but one can choose any number, including all methods. Each method is used to make a prediction of the query job and each method votes for that prediction. The weights obtained from the validation set are used to give more priority to methods that performed better in the validation phase. One could use no weights, but in case of a tie there would be no way to decide which output to choose, and choosing randomly could result in a low accuracy choice.

At prediction time, when a user query arrives, the tool consults the database to convert the query parameters to the format expected by the models. For example, it queries the database to obtain the key of the job's current working directory. Then, the trained models are loaded and predictions are made. After that, each model votes and the final estimate is issued.

TABLE I
DESCRIPTION OF THE FEATURES USED IN THE PREDICTORS.

| Feature | Type | Description |
|---|---|---|
| User ID | Category | User who submitted the job |
| Group ID | Category | User group that submitted the job |
| Queue ID | Category | Number of the queue the job has been submitted to |
| Working directory | Category | Directory where the job executes |
| ResReq | Category | Resources requested (e.g. architecture type, GPU) |
| Command | Category | Command executed |
| Priority | Number | User priority |
| Submission time | Number | Time at which the job was submitted |
| Requested time | Number | Amount of time requested to execute the job |
| Requested processors | Number | Number of processors requested at the submission time |
| Weekday | Number | Day of the week in which the job was submitted |
| Time since midnight | Number | Time of the day at which the job was submitted |

*B. Prediction Techniques*

In this section we describe the machine learning algorithms implemented in our tool. We have used four methods that are further described below: Support Vector Machines (SVMs) with Linear and Radial Basis Function (RBF) kernels (*svm-1* and *svm-2*), Random Forests (*rforest*), Multilayer Perceptrons (MLPs) (*mlp-1* and *mlp-2*), and *k* Nearest Neighbors (kNN) (*knn-1* and *knn-2*).

*1) Support Vector Machines:* SVMs [10] are machine learning algorithms originally proposed for two-group classification problems that use kernel methods to map input vectors to (possibly) non-linear high-dimension feature spaces. In our tool we do multi label classification by using a one *versus* all approach. We always maintain two SVM models: (1) one that uses a linear kernel and regularization constant $C = 0.01$, and (2) another that is the result of using cross validation to select the best model from the parameters found in Table II. In the table, sample influence ($\gamma$) refers to the $\gamma$ term in the RBF kernel, defined as $\exp\left(-\gamma \left|x - x'\right|^2\right)$, and determines the influence a single training example has in the optimization process.

TABLE II
PARAMETER SPACE SEARCHED FOR FINDING THE BEST PERFORMING SVM MODEL.

| Kernel | Regularization ($C$) | Sample influence ($\gamma$) |
|---|---|---|
| Linear | 0.1, 1, 10, 100 | — |
| RBF | 0.1, 1, 10, 100 | $1 \times 10^{-3}, 1 \times 10^{-4}$ |

*2) Random Forests:* Random Forests [11] are models that combine the predictions of several randomized decision trees that are built with bootstrap samples from the training set. In our implementation the forests use 20 trees that choose which nodes to split based on the information gain criterion [12].

*3) Multilayer Perceptron:* An MLP is a feed-forward fully-connected neural network. In our tool we built two such networks: (1) a fixed three-layer network with layer sizes 128, 64 and 32 that used Stochastic Gradient Descent (SGD) and (2) a network whose parameters are obtained by means of cross-validation in the training set. The parameters used for cross-validation are shown in Table III. In the table, SGDNM means Stochastic Gradient Descent with Nesterov's Momentum [13, 14]. The $\beta_1$ and $\beta_2$ parameters found in the table are only used by the Adam algorithm [15].

TABLE III
PARAMETER SPACE SEARCHED FOR FINDING THE BEST PERFORMING MLP MODEL.

| | |
|---|---|
| Layer sizes | (256, 128, 64, 32), (128, 64, 32), (256, 64, 32) |
| Optimization algorithm | SGDNM [13], Adam [15], L-BFGS |
| Learning rate update | Constant, inverse scaling |
| Shuffle | True, False |
| Activation function | $\tanh$ |
| Regularization term ($\alpha$) | 0.01, 0.001, 0.0001 |
| First moment decay ($\beta_1$) | 0.9 |
| Second moment decay ($\beta_2$) | 0.999 |

*4) k Nearest Neighbors:* *k* Nearest Neighbors (kNN) is a learning method that uses the *k* closest examples in the training set to make a prediction. In our tool we implemented two modes of operation: (1) regular voting-based classification and (2) a regression-based prediction service [16] that was adapted to return the bin closest to the predicted value. Both models used $k = 5$.

*C. Using the prediction tool*

In this initial implementation the tool is provided as a command-line binary called `lspredict`. To ease the burden on users having to learn a new command-line tool, `lspredict` was implemented to support the same syntax of `bsub`, the LSF command used for job submission. Hence, to make a prediction, the user just has to prepend `lspredict` to her job submission command.

Internally, `lspredict` takes the user-provided command and modifies it to include the `-H` flag, which makes the job transition to a suspended state until explicitly resumed by a user. After this fake job is submitted, `lspredict` fetches the job's representation from LSF, which is then used to make a prediction. After a prediction is made, the fake job is killed.

## III. EVALUATION

In this section, we describe the setup to evaluate the performance of the proposed tool; i.e. to predict job memory usage when users submit their jobs to a cluster. For this, we used data from two real production systems. Moreover, we present the results compared to a base line performance, that is the mode of the training set. Finally, we discuss why the combination of predictors may be the best option for predicting memory usage.

### A. Environment Setup

We used two systems to evaluate our tool. One of them is a 26-node POWER8 cluster, used by IBM Research division, and the other is a production system composed of 2,128 x86 nodes.

The jobs in these systems are a mix of multi- and single-host applications. The applications from the POWER8 cluster have various characteristics as they come from different areas such as big data, cognitive computing, and more traditional HPC workloads, such as energy production models. The x86 system runs production applications, which are routinely executed by users. We selected these two systems as they represent two types of workloads: one more varied and the other more regular.

We analyzed the performance of the tool with 25,000 jobs of each system, in 5 segments of 5,000 jobs each. For each segment we trained the models with 9,000 jobs and validated them with 1,000 jobs, as previously described. It is important to notice that it is not enough to merely have disjoint sets for training and testing for correct evaluation. We need to consider the temporal relationship between jobs. Maximum memory used is a piece of information that is only available once a job has finished. Therefore the training set must only include finished jobs by the time the current testing job is evaluated. This will naturally occur in production usage of the tool, but care must be taken when evaluating a prediction algorithm using the tool in off-line mode.

It is possible to iterate over the testing set and for each job find all jobs that have finished and use them to train a model. However, doing so may be too expensive, as some of the methods we used may take several minutes to run. We solve this problem by only retraining after 5,000 new jobs have been accumulated. This approach would be convenient to users in production. Since trained models are persisted, new prediction requests can be serviced immediately.

In our evaluation we limited the training/validation set size to the most recently finished 10,000 jobs. We do that because, typically, a heuristic known as "principle of persistence" holds. That is, memory usage tends to persist, making the recent past a good predictor of the near future. Although we fixed the training/validation size to 10,000, one could find a set size that optimizes performance by using some metaheuristic such as Genetic Algorithms.

The traditional way to estimate test error related to fitting a particular machine learning method is to use a validation set. In our tool, we used the validation performance of each method as a weight in the voting scheme. The hope was that (1) the testing error would be similar to the validation error, and (2) combining top methods in the validation would yield good predictions. In our experiments, we confirmed the second hypothesis, even though the first one was not completely correct.

### B. Analysis

In Tables IV and V we compare the performance of the mode and each machine learning method for the validation and test sets during each of the 5,000-job segments. We highlight the top four methods in both sets. One would expect to see a similar pattern in the validation and testing set. However, the top four methods are fairly different in those sets. Particularly, general SVM (svm-2) performs well on validation set, but poorly in the testing data. Meanwhile, the linear SVM model (svm-1), a method with less variance, is much more consistent in the test set, even though it does not make to the top four. This suggests that the SVM (svm-2) may be over-fitted. Moreover, the Random Forests (rforest) method performs well in general, except for a few segments in the x86 system. As a consequence, if we rely only on this method, we may have a worse performance in some situations due to the stochastic nature of the method. Multilayer Perceptron with cross-validation (mlp-2) seems to over-fit whereas the fixed version (mlp-1) appears to generalize better. More importantly, the more sophisticated kNN (knn-2) is not very consistent. This method is used in production in the XSEDE grid to predict queue waiting time. We have implemented this method ourselves for hybrid cloud environments with reasonably good results for both running time and queue waiting time [8]. However, as one can see from Tables IV and V other methods outperformed this form of kNN in several segments. An explanation for this result may be that the kernel method that is applied in this model might be more suitable for regression tasks. The fact that the simple kNN (knn-1) performs better suggests this conclusion.

Even though, some methods are weak, i.e. they do not perform well in the testing data, the polling strategy we used consistently beats the mode, as it can be seen in Figure 2. This strategy appears to reduce the risk of using a single model, since models tend to make mistakes differently. The weak methods do not match up in the voting, only the strong ones. Consequently, the highly voted bins are from strong methods. In order to confirm this hypothesis, we used the accuracy of the testing set as a weight. We cannot do this in production, since this data is only known after the job for which we are predicting memory usage has finished. However, we use these optimal but impossible weights here to compare with the weights we obtained from validation. Figure 3 compares the performance of the polling method with those two sets of weights. As one can see, the method with the validation weights has a similar performance compared with the perfect weights, and sometimes performs even slightly better, though this should be only a result of the stochasticity of the method.

TABLE IV
PREDICTOR PERFORMANCE FOR x86 SYSTEM (TOP 4 IN BOLD).

| | Validation performance | | | | | | | |
|---|---|---|---|---|---|---|---|---|
| segment | mode | svm-1 | svm-2 | rforest | mlp-1 | mlp-2 | knn-1 | knn-2 |
| 0 | 0.7457 | **0.8837** | 0.8769 | **0.9331** | 0.8753 | **0.9014** | 0.8742 | 0.8597 |
| 1 | 0.6987 | 0.9183 | **0.9346** | **0.9588** | 0.9106 | **0.9331** | **0.9313** | 0.9136 |
| 2 | 0.7296 | 0.9047 | **0.9202** | **0.9476** | 0.8934 | **0.9184** | **0.9256** | 0.9049 |
| 3 | 0.6466 | 0.9139 | **0.9194** | **0.9477** | 0.8974 | **0.9214** | **0.9233** | 0.8961 |
| 4 | 0.6121 | 0.8144 | 0.8209 | **0.8924** | 0.8009 | **0.8246** | **0.8657** | **0.8298** |
| | Test performance | | | | | | | |
| segment | mode | svm-1 | svm-2 | rforest | mlp-1 | mlp-2 | knn-1 | knn-2 |
| 0 | 0.7910 | **0.8606** | 0.3948 | 0.6546 | **0.8610** | **0.7278** | 0.6558 | **0.6826** |
| 1 | 0.6024 | **0.7448** | 0.1202 | **0.7544** | **0.7448** | 0.7180 | **0.7540** | 0.5492 |
| 2 | 0.6626 | **0.8640** | 0.0484 | **0.8698** | **0.8630** | **0.8666** | 0.6778 | 0.6770 |
| 3 | 0.8066 | **0.8836** | 0.2464 | **0.8918** | 0.7726 | **0.8842** | **0.8792** | 0.5166 |
| 4 | 0.7742 | **0.8038** | 0.7568 | 0.7812 | **0.8014** | **0.8140** | 0.7772 | **0.8034** |

TABLE V
PREDICTOR PERFORMANCE FOR POWER8 SYSTEM (TOP 4 IN BOLD).

| | Validation performance | | | | | | | |
|---|---|---|---|---|---|---|---|---|
| segment | mode | svm-1 | svm-2 | rforest | mlp-1 | mlp-2 | knn-1 | knn-2 |
| 0 | 0.9154 | 0.9509 | **0.9528** | **0.9996** | 0.9187 | **0.9783** | **0.9624** | 0.9254 |
| 1 | 0.9464 | 0.9641 | **0.9642** | **0.9992** | 0.9460 | **0.9839** | **0.9720** | 0.9464 |
| 2 | 0.9663 | 0.9704 | **0.9740** | **0.9994** | 0.9663 | **0.9812** | **0.9746** | 0.9603 |
| 3 | 0.9439 | 0.9537 | **0.9543** | **0.9996** | 0.9439 | **0.9789** | **0.9620** | 0.9434 |
| 4 | 0.8434 | 0.8767 | **0.8786** | **0.9982** | 0.8459 | **0.9296** | **0.9020** | 0.8414 |
| | Test performance | | | | | | | |
| segment | mode | svm-1 | svm-2 | rforest | mlp-1 | mlp-2 | knn-1 | knn-2 |
| 0 | 0.9856 | **0.9856** | 0.0024 | **0.9890** | **0.9848** | 0.0666 | **0.9796** | 0.9796 |
| 1 | 0.9628 | **0.9610** | 0.0014 | **0.9656** | **0.9610** | 0.7772 | **0.9620** | 0.9610 |
| 2 | 0.9398 | **0.9436** | 0.0042 | **0.9428** | **0.9398** | 0.1322 | 0.2912 | 0.2826 |
| 3 | 0.8276 | **0.8264** | 0.0092 | **0.8360** | 0.8162 | 0.7952 | **0.8278** | **0.8168** |
| 4 | 0.7386 | **0.7412** | 0.0232 | **0.7474** | **0.7400** | 0.5162 | **0.7410** | 0.7274 |

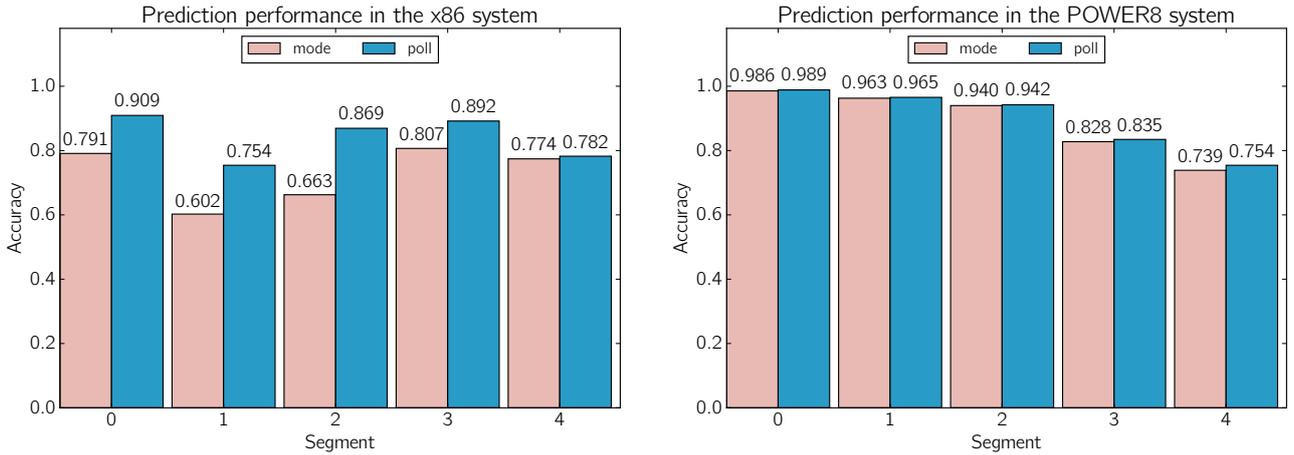

Fig. 2. Comparison between mode and poll for both environments. In the left graph we see the performance in the x86 system, while in the right we see the performance of the POWER8 system.

In Figure 2 we also see that the different characteristics of both clusters play an important role in the predictors'

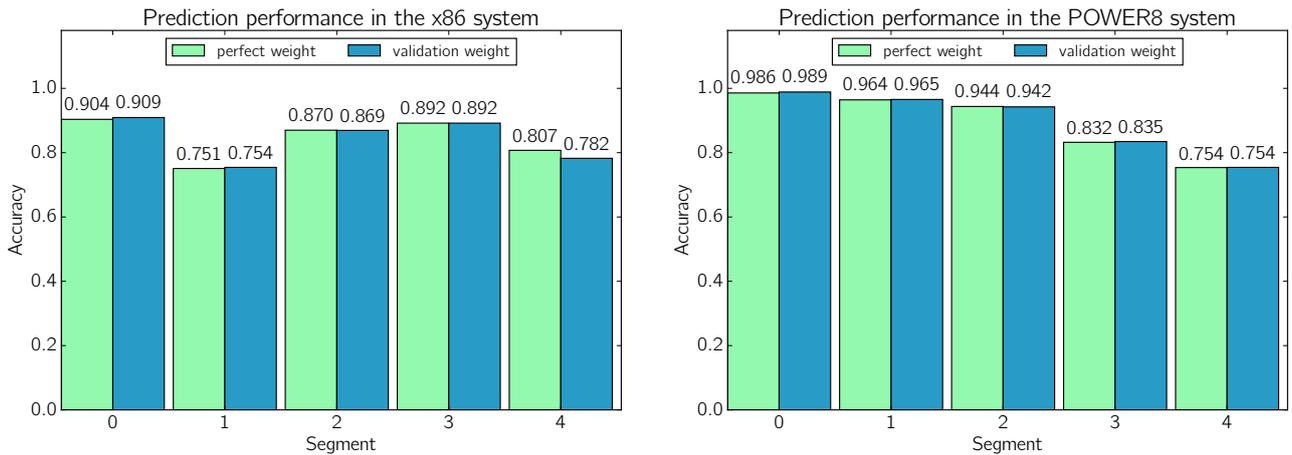

Fig. 3. Comparison between poll using perfect weight (i.e. weights obtained from the accuracy of the test set) and poll using validation weights for both environments. In the left graph we see the performance in the x86 system, while in the right we see the performance of the POWER8 system.

TABLE VI
PREDICTOR PERFORMANCE OF THE TOP $n$ ESTIMATORS FOR THE X86 SYSTEM.

| Segment | Top 1 | Top 2 | Top 3 | Top 4 | Top 5 | Top 6 | All |
| --- | --- | --- | --- | --- | --- | --- | --- |
| 0 | 0.6546 | 0.6546 | 0.9088 | 0.9092 | 0.9092 | 0.9092 | 0.9092 |
| 1 | 0.7544 | 0.7544 | 0.7538 | 0.7542 | 0.7542 | 0.7546 | 0.7546 |
| 2 | 0.8698 | 0.8698 | 0.7018 | 0.8692 | 0.7250 | 0.8690 | 0.8690 |
| 3 | 0.8918 | 0.8918 | 0.8918 | 0.8918 | 0.8918 | 0.8844 | 0.8842 |
| 4 | 0.7812 | 0.7812 | 0.7774 | 0.7824 | 0.7822 | 0.8152 | 0.8092 |

TABLE VII
PREDICTOR PERFORMANCE OF THE TOP $n$ ESTIMATORS FOR THE POWER8 SYSTEM.

| Segment | Top 1 | Top 2 | Top 3 | Top 4 | Top 5 | Top 6 | All |
| --- | --- | --- | --- | --- | --- | --- | --- |
| 0 | 0.9890 | 0.9890 | 0.9890 | 0.9886 | 0.9886 | 0.9878 | 0.9868 |
| 1 | 0.9656 | 0.9656 | 0.9656 | 0.9654 | 0.9652 | 0.9644 | 0.9642 |
| 2 | 0.9428 | 0.9428 | 0.9428 | 0.9424 | 0.9426 | 0.9426 | 0.9426 |
| 3 | 0.8360 | 0.8360 | 0.8360 | 0.8346 | 0.8360 | 0.8328 | 0.8316 |
| 4 | 0.7474 | 0.7474 | 0.7474 | 0.7542 | 0.7530 | 0.7546 | 0.7534 |

performance: in the POWER8 system, memory usage is highly concentrated in a single bin, as can be seen by the good performance exhibited by a predictor based on the mode. Therefore, since the data set is varied, models fail to detect a useful structure, and do not have much room for improvement over the mode. In the x86 system, which runs production code, models seem to be able to learn some structure and can greatly outperform the mode.

Tables VI and VII display how the performance of the polling-based method varies as the number of voters increase. One can see that a polling-based approach does reduce the risk of relying on a single predictor. In segment 0 of the x86 system, for example, using three models outperforms the best predictor in approximately 25%. This improvement does not come without a cost, though, as adding more voters reduces performance slightly.

## IV. RELATED WORK

Research efforts related to our work are mostly in the area of predicting two types of features: execution time and queue waiting time. Some of these efforts can be used to predict other features. Here is a brief description of some of these efforts.

### A. Queue time predictions

Estimating how long a job will wait in a queue before its execution is a key component for deciding where jobs should be placed. There are several techniques available in the literature. For example, Li et al. [17] investigated methods and algorithms to improve queue wait time predictions. Their work assumes that similar jobs under similar resource states have similar waiting times as long as the scheduling policy and its configuration remains unchanged for a considerable amount of time.

Nurmi et al. [18] introduced an on-line method/system, known as QBETS, for predicting batch-queue delay. Their main motivation is that job wait times have variations that make it difficult for end-users to plan themselves and be productive. The method consists of three components: a percentile estimator, a change-point detector, and a clustering procedure. The clustering procedure identifies jobs of similar characteristics; the change-point detector determines periods of stationarity for the jobs; and the percentile estimator calculates a quantile that serves as a bound on future wait time.

Kumar and Vadhiyar [19] developed a technique that defines which jobs can be classified as *quick starters*. These are jobs with short waiting times compared to the other jobs waiting for resources. Their technique considers both job characteristics such as request size and estimated runtime, and the state of the system, including queue and processor occupancy states.

More recently, Murali and Vadhiyar [20] proposed a framework called Qespera for prediction of queue waiting times for HPC settings. The proposed framework is based on clustering using history of job submissions and executions. The weights associated with the features for each prediction are adapted depending on the characteristics of the target and history jobs.

### B. Runtime predictions

Smith [16] developed a method/system for estimating both queue wait time and job runtime. The method is based on IBL (Instance-Based Learning) techniques and leverages genetic algorithms (GA) to refine input parameters of the method. This system is used by XSEDE to predict queue wait time.

Yang et al. [21] proposed a technique to predict the execution time of jobs in multiple platforms. Their method is based on data collected from short executions of a job and the relative performance of each platform.

Tsafrir et al. [3] developed a technique for scheduling jobs based on system-generated job runtime estimates, instead of using user provided estimates. For the runtime estimates, they analyzed several workloads from supercomputer centers and found out that users tend to submit similar jobs over a short period of time. Therefore, their estimations are based on the average time of the previous two actual job runtime values. Gaussier et al. [22] studied the impact of using machine learning for run time predictions on job schedulers based on backfilling.

### C. Memory usage predictions

Efforts on memory usage predictions rely mostly on benchmarking rather than on scheduler logs [21, 23, 24, 25, 26, 27]. For instance, Matsunaga and Fortes [26] assessed various machine learning techniques for predicting spatio-temporal utilization of resources by user applications; memory is one of the resources investigated. Their experiments focused on two bioinformatics applications and involved the execution of the applications with different parameter configurations.

Another example is from Wood et al. [27] who designed an approach for estimating the resource requirements of user applications motivated by the need to move such applications to virtualized environments. Their approach relies on a set of microbenchmarks to profile the different types of virtualization overhead on a given platform and a regression-based model to map the native system usage profile into a virtualized one.

We built on top of existing efforts in the literature to help users have predictions for a variety of scenarios. In particular usability and how to handle inaccuracy for these predictions are essential for adoption of this kind of technology. This paper focuses on using batch scheduler logs to predict job memory usage, but such predictions can be used for other types of resources.

## V. CONCLUSION

HPC users still have difficulties to specify resource requirements to run their applications and the HPC community always welcomes novel techniques and tools to help users make better use of cluster environments. This type of tools is even more important in the context of HPC Cloud [28]. In this scenario, we introduced here a tool aimed at predicting memory requirements of jobs submitted to cluster environments. We describe in details the tool and how it was incorporated into the LSF batch scheduler.

The proposed prediction tool relies on machine learning methods to make predictions. From our experiments using job traces from two production environments, we observed that there is no single machine learning method that produces the best predictions. Therefore, the proposed tool leverages the predictions of all methods and select the most promising ones at a given situation.

### ACKNOWLEDGEMENTS

This work has been partially supported by FINEP/MCTI under grant no. 03.14.0062.00.